\documentstyle[aps,prl,multicol,epsf,graphicx,amssymb,amsfonts]{revtex}

\begin{document}
\title{Universal renormalization-group dynamics at the onset of chaos in logistic maps and
nonextensive statistical mechanics}
\author{F. Baldovin$^{1}$ and A. Robledo$^{2}$
\thanks{E-mail addresses: baldovin@cbpf.br,
robledo@fenix.ifisicacu.unam.mx}}
\address{$^{1}$Centro Brasileiro de Pesquisas F\'{i}sicas, 
Rua Xavier Sigaud 150,\\
22290-180 Rio de Janeiro-RJ, Brazil. \\
$^{2}$Instituto de F\'{i}sica,\\
Universidad Nacional Aut\'{o}noma de M\'{e}xico,\\
Apartado Postal 20-364, M\'{e}xico 01000 D.F., Mexico.}
\maketitle

\begin{abstract}
We uncover the dynamics at the chaos threshold $\mu _{\infty }$ of the
logistic map and find it consists of trajectories made of intertwined power
laws that reproduce the entire period-doubling cascade that occurs for $\mu
<\mu _{\infty }$. We corroborate this structure analytically via the
Feigenbaum renormalization group (RG) transformation and find that the
sensitivity to initial conditions has precisely the form of a $q$-exponential, 
of which we determine the $q$-index and the $q$-generalized
Lyapunov coefficient $\lambda _{q}$. Our results are an unequivocal validation
of the applicability of the non-extensive generalization of Boltzmann-Gibbs
(BG) statistical mechanics to critical points of nonlinear maps.
\end{abstract}


\begin{multicols}{2}
Critical points of nonlinear maps offer a suitable playground for
testing the validity of the non-extensive generalization of the
Botzmann-Gibbs (BG) statistical mechanics proposed by Tsallis over a decade
ago \cite{tsallis0,tsallis1}. Here we describe universal properties
related to the dynamics of iterates at the onset of chaos in unimodal maps 
\cite{schuster1}, that provide a literal confirmation of the generalized
non-extensive theory. To this end we employ the celebrated one-dimensional
logistic map, $f_{\mu }(x)=1-\mu \left| x\right| ^{2}$,$\;-1\leq x\leq 1$,
and the properties of its renormalization group (RG) fixed point, to present
evidence of previously unexposed scaling properties at the onset of chaos 
$\mu =\mu _{\infty }$. At this state, the most prominent of the map critical
points, the trajectories of the iterates exhibit an intricate structure,
that we describe and show is governed by the Feigenbaum's RG transformation 
\cite{schuster1}.

The domain of validity of BG statistical mechanics has been implicitly
challenged by the proposal of its non-extensive generalization. Subsequent
studies have offered experimental and numerical evidence that point out both
the inadequacy of the standard BG statistics and the plausible competence of
the generalized theory in describing various types of phenomena and systems.
This theoretical development represents an exceptional event in the long and
trustworthy history of BG statistical mechanics. However, it is still in the
process of being converted into a rigorously corroborated and fully
understood fact. The suggested circumstances under which the generalized
theory is believed to be applicable, at least with regards to non-linear
dynamical systems, are those associated to a phase space with power-law
sensitivity to initial conditions, to the consequent vanishing of the
largest Lyapunov exponent, and to a fractal, or multifractal geometrical
structure \cite{tsallis1}. Here we show that our results for the dynamics at
the onset of chaos in unimodal maps constitute an unequivocal proof of the
universal validity of the non-extensive statistics at such critical points.

In fact, at the chaos threshold (as well as at other critical points of the map)
the Lyapunov exponent $\lambda_{1}$ 
vanishes, and the sensitivity to initial conditions $\xi _{t}$,
for large iteration time $t$, ceases to obey exponential behavior, exhibiting
instead power-law behavior \cite{oldrefs1}. In order to describe the
dynamics at such critical points, the $q$-exponential expression 
\begin{equation}
\xi _{t}=\exp _{q}(\lambda _{q}t)\equiv [1-(q-1)\lambda _{q}t]^{-1/(q-1)},
\end{equation}
containing a $q$-generalized Lyapunov coefficient $\lambda _{q}$, has been proposed 
\cite{tsallis2}. This expression is based on the non-extensive entropy of
Tsallis \cite{tsallis1}. As a companion to this, generalizations for the
Kolmogorov-Sinai (KS) entropy $K_{q}$ and for the Pesin identity $\lambda
_{q}=$ $K_{q}$, $\lambda _{q}>0$ have also been introduced \cite{tsallis2}
(the standard expressions are recovered when $q\rightarrow 1$). Several
recent studies \cite{tsallis2,tsallis3,tsallis4,lyra1}, that probed numerically the
onset of chaos of the logistic map and its generalization to nonlinearity 
$\zeta>1$ 
($f_{\mu,\zeta}(x)\equiv 1-\mu \left| x\right| ^{\zeta}$,$\;-1\leq x\leq 1$), 
have revealed a series of precise connections between the Tsallis
entropic index $q$ and the map basic parameters. Here we present RG
analytical results that corroborate the previously known value of $q$ at 
$\mu _{\infty }$ for $\zeta =2$ and also determine $\lambda _{q}$ for the
first time.

To state our results more precisely, we recollect the following properties.
The logistic map exhibits several types of infinite sets of critical points
that appear as its control parameter $\mu $ varies; 
these correspond, amongst others, to 
period-doubling and chaotic-band-splitting transitions \cite{schuster1}. 
The accumulation point of the
period doublings and also of the band splittings is the Feigenbaum attractor
that marks the threshold between periodic and chaotic orbits, at $\mu
_{\infty }=1.40115...\;$. 
The locations of period doublings (at $\mu =\mu_{n}<\mu _{\infty }$) 
and band splittings (at $\mu =\widehat{\mu }_{n}>\mu_{\infty }$) 
obey, for large $n$, power laws of the form $\mu _{n}-\mu
_{\infty }\sim \delta ^{-n}$ and $\mu _{\infty }-\widehat{\mu }_{n}\sim
\delta ^{-n}$, where $\delta =4.6692...$ is one of the two Feigenbaum's
universal constants. 
For our use below, we recall also the sequence of
parameter values $\overline{\mu }_{n}$ employed to define the diameters 
of the bifurcation forks $d_{n}$ that form the period-doubling cascade
sequence. At $\mu =$ $\overline{\mu }_{n}$ the map displays a `superstable'
periodic orbit of length $2^{n}$ that contains the point $x=0$. For large $n$,
the distances to $x=0$ of the iterate positions in such 
$2^{n}$-cycle that are closest to $x=0$, 
$d_{n}\equiv f_{\overline{\mu }_{n}}^{(2^{n-1})}(0)$, 
have constant ratios: $d_{n}/d_{n+1}=-\alpha $, where $\alpha =2.50290...$ is
the second of the Feigenbaum's constants. A set of diameters with scaling
properties similar to those of $d_{n}$ can also be defined for the band
splitting sequence \cite{schuster1}. For clarity of presentation of our
results we shall only use absolute values of positions, so that the dynamics
of iterates do not carry information on the self-similar properties of
``left'' and ``right'' symbolic dynamic sequences \cite{schuster1}. This
choice does not affect results on the sensitivity to initial conditions.
Below, $d_{n}$ means $\left| d_{n}\right| $.

The main points in the following analysis are: 1) The iterates {\it at} $\mu
_{\infty }$ follow trajectories that proceed in a concerted manner according
to the entire period-doubling cascade that takes place for $\mu <\mu
_{\infty }$. The positions of the trajectories are given in fact in terms 
of the diameters $d_{n}(\overline{\mu }_{n})$ of the $2^{n}$-supercycles. 
2) As a consequence, the sensitivity to initial
conditions also evolves in agreement to the period-doubling cascade. 3) The
bounds, or envelops, as well as other monotonic subsequences, of both a
single trajectory $x_{t}$ and of the sensitivity to initial conditions $\xi _{t}$ 
have precisely the form of a $q$-exponential. For $\xi_t$ we have 
$q=1-\ln 2/\ln \alpha $ and $\lambda _{q}=\ln \alpha /\ln 2$. 4) These
results are obtainable via the fixed-point solution $g(x)$ of the RG
doubling transformation, consisting of functional composition and rescaling: 
${\bf R}f(x)\equiv \alpha f(f(x/\alpha )$.

\begin{figure}
\begin{center}
\includegraphics[width=8cm,angle=0]{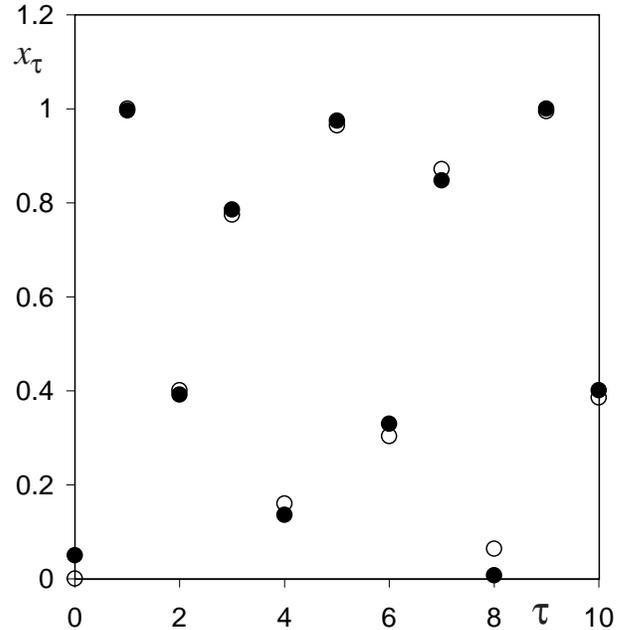}
\end{center}
\caption{\small 
Absolute values of positions of the first $10$
iterations $\tau $ for two trajectories of the logistic map with
initial conditions $x_{0}=0$ (empty circles) and $x_{0}=\delta
\simeq 5\;10^{-2}$ (full circles). Plotted quantities are dimensionless.} 
\end{figure}

To begin, we show in Fig. 1 the absolute values of the positions of two
trajectories of the logistic map with initial conditions $x_{0}=0$ and 
$x_{0}=\delta \simeq 5\;10^{-2}$, for the first $10$ iterations $\tau $. For 
$\tau =1$ the positions are $x_{1}=1$ and $x_{1}\simeq 1-3.5\;10^{-3}$, and it can
be observed that the difference between the two positions at times 
$\tau =2$, $4$, and $8$ grows progressively. In Fig. 2 we show 
the same first
trajectory $x_{0}=0$ and a second one $x_{0}=\delta\simeq10^{-4}$, 
up to $\tau=1000$. 
In the logarithmic scales 
it can be clearly appreciated that they
consist of interwoven monotonic position subsequences with power-law decay.
We are interested in the position subsequences that are generated by the
time subsequences $\tau =2^{n}+2^{n-k}$, $k=0,1,...\;$. As we show below, the
values for the trajectory subsequence $\tau =2^{n}$, with $x_{0}=0$, are
asymptotically given by $x_{\tau }=d_{n}$, $n\geq 0$. More generally, each
position subsequence $\tau =2^{n}+2^{n-k}$, obtained with $x_{0}=0$, is given
by $g^{(2^{k}+1)}(0)\;d_{n}^{k}$, where $d_{n}^{k}\equiv 
f_{\overline{\mu }_{n}}^{(2^{n-k-1})}(0)$ and $n\geq k$. 
The itinerary of the iterate starting
at $x_{0}=0$ can be clearly observed in Fig. 2. The position approaches the
origin $x_{0}=0$ progressively as $n$ increases every time that $\tau =2^n$, 
but in between the values $2^{n}$ and $2^{n+1}$ it returns in an
oscillatory manner towards $x_{1}=1$, repeating twice the positions visited
in the previous cycle between $2^{n-1}$ and $2^{n}$ and introducing a new
position between these two sub-cycles. For small $n$ the positions are
approximately repeated, but they become accurately reproduced as $n$
increases. The whole time series has the period-doubling structure.

\begin{figure}
\begin{center}
\includegraphics[width=8cm,angle=0]{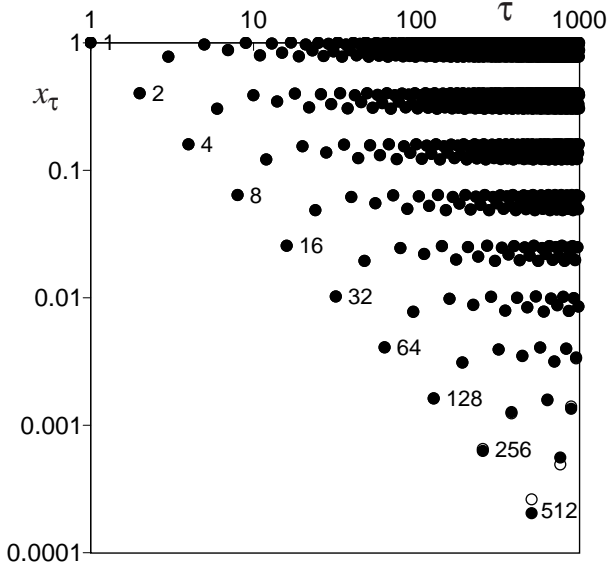}
\end{center}
\caption{\small 
Absolute values of positions of the first $1000$ 
iterations $\tau $ for two trajectories of the 
logistic map with initial conditions $x_{0}=0$ (empty circles) 
and $x_{0}=\delta =10^{-4}$ (full circles) in
logarithmic scales. The power-law decay of several time 
subsequences can be
clearly appreciated.} 
\end{figure}

To interpret the dynamics in Figs. 1 and 2 in terms of the RG transformation,
we consider ${\bf R}$ applied $n$ times to the fixed-point map $g(x)$, i.e.: 
\begin{equation}
g(x)={\bf R}^{(n)}g(x)\equiv \alpha ^{n}g^{(2^{n})}(x/\alpha ^{n}).
\end{equation}
We determine the trajectory positions at times $\tau =2^{n}$, with $x_{0}=0$.
Since $g(0)=1$, we have $g^{(2^{n})}(0)=$ $\alpha ^{-n}$ and because 
$d_{n}/d_{n+1}=\alpha $ with $d_{0}=1$ implies $d_{n}=$ $\alpha ^{-n}$, we
also have $d_{n}=g^{(2^{n})}(0)$. Thus, we obtain the diameters $d_{n}$ to be
the positions $x_{2^{n}}$. This result can be expressed as a $q$-exponential
if we shift the time variable by one unit, $t=2^{n}-1$, and rearrange 
$\alpha ^{-n}$ as $(1+t)^{-\ln \alpha /\ln 2}$. We obtain 
\begin{equation}
x_{t}=\exp _{Q}(\Lambda _{Q}t),
\end{equation}
with $Q=1+\ln 2/\ln \alpha $ and $\Lambda _{Q}=-\ln \alpha /\ln 2$. Other
position subsequences $\tau =2^{n}+2^{n-k}$ can be put in the form of 
$q$-exponentials with the same values of $Q$ and $\Lambda _{Q}$.

The expression for the sensitivity to initial conditions can be derived with
the use of the following approximate property (that becomes asymptotically
exact in the limit $n\to\infty$) 
\begin{equation}
g^{(2^{n})}(d_{j})=\frac{1}{\alpha ^{n}}-\frac{\mu _{\infty }}
{\alpha ^{2j-n}}=d_{n}-\mu _{\infty }d_{2j-n},\ n\leq j.  \label{key}
\end{equation}
To prove the above, note first that 
$g^{(2^{n})}(d_{j})=g^{(2^{n}+2^{j})}(0)=\alpha ^{-n}g^{(2^{j-n}+1)}(0)$, 
or, since $g^{(2^{j-n}+1)}(0)=$ $g(d_{j-n})$, 
$g^{(2^{n})}(d_{j})=$ $\alpha ^{-n}g(d_{j-n})$. The preceding equality,
together with $g(d_{j-n})=1-$ $\mu _{\infty }d_{j-n}^{2}=1-\mu _{\infty
}\alpha ^{2n-2j}$, yields Eq. (\ref{key}). Now, the distance between
positions $x_{2^{n}}(d_{j})=g^{(2^{n})}(d_{j})$ and 
$x_{2^{n}}(d_{i})=g^{(2^{n})}(d_{i})$ at time $\tau =2^{n}$ can be written
with the use of Eq. (\ref{key}) as $x_{2^{n}}(d_{j})-x_{2^{n}}(d_{i})=$ 
$[x_{2^{0}}(d_{j})-x_{2^{0}}(d_{i})]\alpha ^{n}$, or with use of the shifted
time variable $t=2^{n}-1$ as 
\begin{equation}
x_{t}(d_{j})-x_{t}(d_{i})=(x_{0}(d_{j})-x_{0}(d_{i}))\alpha ^{n}.
\end{equation}
The sensitivity to initial conditions $\xi _{t}$ is defined as 
\begin{equation}
\xi _{t}\equiv \lim_{\left| \Delta
x_{0}\right| \to 0}\left| x_{t}(d_{j})-x_{t}(d_{i})\right| /\left|
x_{0}(d_{j})-x_{0}(d_{i})\right| 
\end{equation}
(where $\lim_{\left| \Delta x_{0}\right|\to 0}$ is equivalent to 
$\lim_{i,j\to \infty ,i\neq j}$). 
$\xi_t$ can then be written,
considering that $\alpha ^{n}=$ $(1+t)^{\ln \alpha /\ln 2}$, as the 
$q$-exponential 
\begin{equation}
\xi _{t}=\exp _{q}(\lambda _{q}t),
\end{equation}
where $q=1-\ln 2/\ln \alpha $ and $\lambda _{q}=\ln \alpha /\ln 2$. Notice
that $q=2-Q$ as $\exp _{q}(y)=1/\exp _{Q}(-y)$. 
The previous construction
applies strictly to initial positions that lie on the attractor;
nevertheless, we remark that all other positions tend to the attractor with a
power-law behavior (see \cite{robledo0,baldovin1}), so that after an initial
transient their dynamics becomes practically indistinguishable from the
situation we describe here. 
In Fig. 3 we show the $q$-logarithm of $\xi _{t}$
vs $t$ (with $q=1-\ln 2/\ln \alpha =0.2445...$), from a numerical simulation
of two trajectories with initial conditions $x_{0}=0$ and 
$x_{0}=\delta\simeq10^{-8}$. 
The result is a straight line with slope very close to $\lambda
_{q}=\ln \alpha /\ln 2=1.3236...\;$. This corroborates the RG prediction 
(the $q$-logarithm, $\ln _{q}y\equiv (y^{1-q}-1)/(1-q)$, is the inverse of $\exp
_{q}(y)$).

\begin{figure}
\begin{center}
\includegraphics[width=8cm,angle=0]{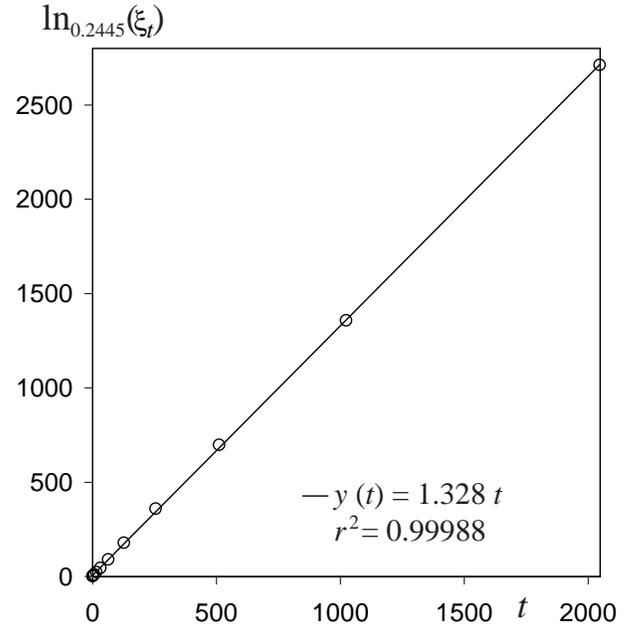}
\end{center}
\caption{\small 
The $q$-logarithm of sensitivity to initial 
conditions $\xi_{t}$ vs $t$, with 
$q=1-\ln 2/\ln \alpha =0.2445...$ and initial conditions 
$x_{0}=0$ and $x_{0}=\delta \simeq 10^{-8}$ (circles). 
The full line is the
linear regression $y(t)$. 
As required, the numerical results reproduce a
straight line with a slope very close to $\lambda _{q}=\ln \alpha /\ln
2=1.3236...\;$.} 
\end{figure}

Interestingly, both $x_{t}/x_{0}$ and $\xi _{t}$ can be seen to satisfy the
dynamical fixed-point relations $h(t)=\alpha h(h(t/\alpha ))$ with $\alpha
=2^{1/(Q-1)}$ and $\alpha =2^{1/(q-1)}$, respectively. In relation to this,
we note that the static fixed-point solution $f^{*}(x)/x$ to the Feigenbaum
RG recursion relation for the case of the tangent bifurcation, obtained by
Hu and Rudnick \cite{schuster1,hu1} to study the intermittency
transition in the $\zeta$-logistic map, 
has the form of a $q$-exponential with 
$q=2$. This has been pointed out recently \cite{robledo0,baldovin1}, 
where in addition
it has been shown that this solution applies too, but now with $q=3$,
to the period-doubling transitions that take place at $\mu _{n}<\mu _{\infty
}$. We recall that, for the transition to periodicity of order $n$, the RG
transformation is applied to the $n$-th composition $f^{(n)}$ of the
original map in the neighborhood of one of the $n$ points tangent to the
line with unit slope, and a shift is made of the origin of coordinates to
that point. The RG fixed-point map is $f^{*}(x)=x\exp _{q}(ux^{q-1})$, where 
$u$ is the leading expansion coefficient of $f^{(n)}$ and where the
recursion relation is satisfied with $\alpha =2^{1/(q-1)}$. In the
neighborhood of the intermittency and period-doubling transitions the time
evolution of iterates follow monotonic paths set by the form of the map
itself, and are also $q$-exponentials. For these types of critical points
the static and dynamic properties are simply related and obey expressions of
the same form \cite{robledo0,baldovin1}. Not only the entropic index $q$ can be
plainly identified, but the $q$-generalized Lyapunov
coefficient $\lambda _{q}$ turns out to be given by the expansion coefficient 
$u$. Unequivocal corroboration of these results has been obtained recently 
\cite{baldovin1}. 

At the onset of chaos both static and dynamical properties
are more complex. The celebrated RG fixed-point map static solution is
obtained as a power series of a smooth unimodal trascendental function, the
universal $n\rightarrow \infty $ limit of 
$(-\alpha )^{n}f_{\overline{\mu }_{n+1}}^{(2^{n})}(x/(-\alpha )^{n})$; 
likewise, the fractal dimension of the
attractor $d_{f}=0.538045143...$ is obtained considering also the same 
$n\rightarrow \infty $ limit on positions of the $2^{n}$-cycles 
\cite{schuster1,christiansen1}. 
As we have seen here, the multifractal attractor at $\mu
_{\infty }$ imprints an involved structure into the time evolution of the
iterates, that can be resolved in terms of simpler monotonic time
subsequences. Remarkably, these subsequences and the sensitivity to initial
conditions $\xi _{t}$ are analytically reproduced by the same RG
transformation originally applied to describe static properties. These
quantities evolve as universal $q$-exponentials, with $q$ and $\lambda _{q}$
simply expressed in terms of $\alpha $. 
We observe then a connection between dynamic
properties of a strange attractor at the edge of chaos, such as the
$q$-generalized Lyapunov coefficient, and the static properties, described by the
set of distances $d_{n}^{k}$ that make up this multifractal;
this is in the spririt of the Kaplan-Yorke
conjecture \cite{schuster1}.

Under some conditions, exemplified here by critical points in nonlinear
maps, the Lyapunov exponents of a system that measure the strength of phase
space mixing vanish. When this happens dynamic processes become sluggish in
exploring their permissible configurations and may be capable of covering
only a small fraction of the available phase space, even in the limit 
$t\rightarrow \infty $. This fraction may have a fractal dimension smaller
than the total dimension of phase space. These are thought to be the
conditions for failure of BG statistics and applicability of its
nonextensive generalization \cite{tsallis1}, and the significance of our
analytically-backed results with no approximations is a contribution towards
the clarification of this issue. At the onset of chaos of unimodal maps the
reduced phase subspace is represented by the strange attractor, a Cantor
subset of the interval $-1\leq x\leq 1$. It is important to point out that
in this case the permissible positions (configurations) are asymptotically
confined by the attractor and this acts as an inescapable barrier to
movement to other locations. By construction, the dynamics at the onset of
chaos, as well as those restricted to the neighborhood of the intermittency
transitions, describe a purely nonextensive regime. It should be mentioned
that our analysis does not consider the access of trajectories to an
adjacent or neighboring chaotic region, as in the setting of Refs. 
\cite{grigolini1,grigolini2} or that in conservative maps \cite{baldovin2}. 
Hence there is no feedback vehicle for a crossover from a nonextensive
regime with vanishing ordinary Lyapunov exponent to an extensive regime with
a positive one at some $t_{cross}$ as $t\rightarrow \infty $. 

Clearly, our findings have wide-ranging validity, as they apply
to all 
dissipative systems in the Feigenbaum universality class. It
might be possible 
to observe analogous behavior at chaos boundary criticalities in
other classes 
of dynamical systems, such as period doubling in bimodal 
\cite{mackay1} and other 
multiparameter maps 
\cite{kuznetsov1}, and quasiperiodicity in
dissipative systems \cite{feigenbaum1}.
More generally, the properties discussed here are likely to hold strictly for other
types of systems or situations that possess equivalent phase space
limitations. Those systems for which experimental and numerical evidence has
accumulated on BG statistics inadequacy and nonextensive statistics
competency \cite{tsallis1} warrant examination. The more that is learned on
mechanisms and circumstances leading to a hindered phase space the clearer
the physical understanding of the applicability of the nonextensive theory
will become.

\section*{Acknowledgments} 
We would like to thank C. Tsallis,  E.M.F. Curado 
and L.G. Moyano for
useful discussions and comments. AR gratefully acknowledges the hospitality
of the Centro Brasileiro de Pesquisas Fisicas where this work was carried
out and the financial support given by the CNPq processo 300894/01-5
(Brazil). AR was also partially supported by CONACyT grant 34572-E and by
DGAPA-UNAM grant IN110100 (Mexican agencies). FB has benefitted from
partial support by CAPES, PRONEX, CNPq, and FAPERJ (Brazilian agencies).

\end{multicols}

\end{document}